\documentclass[manuscript,screen]{acmart}

\usepackage{hyperref}
\usepackage[hyphenbreaks]{breakurl}

\AtBeginDocument{%
  \providecommand\BibTeX{{%
    \normalfont B\kern-0.5em{\scshape i\kern-0.25em b}\kern-0.8em\TeX}}}
    
\setcopyright{iw3c2w3}
\copyrightyear{2021}
\acmYear{2021}
\acmDOI{}

\acmConference[IR4Children '21]{IR for Children 2000-2020: Where Are We Now? -- Workshop co-located with the 44$^{th}$ International ACM SIGIR Conference on Research and Development in Information Retrieval}{July 15, 2021}{Online Event}

\begin{document}

\title[To Infinity and Beyond!]{To Infinity and Beyond! Accessibility is the Future for Kid' Search Engines}

\author{Ashlee Milton}
\email{AshleeMilton@u.boisestate.edu}
\affiliation{%
  \institution{PIReT - Dept. of Computer Science, Boise State University}
  \city{Boise}
  \country{USA}
  \postcode{83725} 
  }
  
\author{Garrett Allen}
\email{GarrettAllen@u.boisestate.edu}
\affiliation{%
  \institution{PIReT - Dept. of Computer Science, Boise State University}
  \city{Boise}
  \country{USA}
  \postcode{83725} 
  }
  
\author{Maria Soledad Pera}
\orcid{0000-0002-2008-9204}
\email{solepera@boisestate.edu}
\affiliation{%
  \institution{PIReT - Dept. of Computer Science, Boise State University}
  \city{Boise}
  \country{USA}
  \postcode{83725} 
  }

\renewcommand{\shortauthors}{Milton, et al.}

\begin{abstract}
Research in the area of search engines for children remains in its infancy. Seminal works have studied how children use mainstream search engines, as well as how to design and evaluate custom search engines explicitly for children. These works, however, tend to take a one-size-fits-all view, treating children as a unit. Nevertheless, even at the same age, children are known to possess and exhibit different capabilities. These differences affect how children access and use search engines. To better serve children, in this vision paper, we spotlight accessibility and discuss why current research on children and search engines does not, but should, focus on this significant matter.
\end{abstract}



\maketitle

\section{"You got a play-date with destiny!": Foundation}

A wide range of users regularly turn to search engines in varying settings, e.g., work, home, and school \cite{haider2019invisible, pewSE, sentance, SEstats}. For decades, the Information Retrieval (\textbf{IR}) community has introduced algorithmic solutions to improve and expand the performance of search engines; it has also studied the people who utilize them. The bulk of existing research works respond to the requirements of ``general'' users who are, for the most part, adult searchers. We could, however, name other ``non-traditional'' user groups, e.g., expert searchers, searchers for whom English is not their first language, and children\footnote{Here, whenever we state children we refer to any person under the age of 18, as per UNICEF's Convention on the Rights of the Child \cite{unicef1989convention} who have unique needs and expectations when using search engines.}, who routinely use search engines, but struggle to best take advantage of them as they are not designed with non-traditional searchers in mind. Children, in particular, have specialized cognitive and physical requirements when it comes to information seeking using search engines \cite{bilal2010mediated}. With children making up almost a quarter of the population and being heavily reliant on search engines to satisfy their information needs \cite{rowlands2008google}, we scope our discussion to encompass \textit{search} \textit{engines} and \textit{children}. 

From a preliminary literature review on Google Scholar and the ACM Digital Library using queries including ``web search children'' and ``search engine children,'' it emerges that research about children and search engines started in the late 1990s. To our knowledge, the earliest publications include those by Dania Bilal \cite{bilal1998children, bilal1999web}, who aimed to understand how children interact with search engines specifically designed for them, as well as Martyn \citet{wild1996researching}, who studied how children use the Internet for information seeking. These contributions provided the groundwork for future research looking into (i) how children use search engines, (ii) modifying search engines algorithmically for children, and (iii) the design of search engines tailored to children \cite{gossen2016search,danovitch2019growing}. Thus far, promising research in this area has probed each stage of the information seeking process. Given that children struggle with query formulation, it is not unexpected to see projects studying how children express their information needs in the form of queries or introducing new strategies to ease the burden of query formulation \cite{bilal2018children,gupta2013algorithm,vanderschantz2017study}. Query suggestions could, to a degree, facilitate the task of identifying effective queries to initiate the search process. Oftentimes, however, suggested queries are the result of query-log-based strategies which are heavily influenced by searches conducted on search engines by general users. To address this limitation, researchers have produced query suggestion strategies that are more appealing to children and that best respond to their search intent \cite{madrazo2018looking,dragovic2016sven,shaikh2015suggesting}. When looking at results retrieved in response to children's queries, \citet{anuyah2019classroom} and \citet{bilal2019readability} investigate the reading level of SERP and whether they align with a child's reading level. When considering the relevance of retrieved resources, children tend to explore search engine result pages (\textbf{SERP}) sequentially, i.e., from top to bottom \cite{gwizdka2017analysis,landoni2019sonny}, as opposed to deciding relevance through scanning SERP snippets and clicking on a given result. This prompted researchers to investigate ways to both filter and re-rank resources so that those more relevant and suitable for children make their way to the top of the SERP \cite{eickhoff2010suitability,vanderschantz2017internet,patel2016kids, milton2020korsce, bilal2017towards,figueiredo2019false,collins2011personalizing}. From the standpoint of how children use search engines, \citet{foss2012children} as well as \citet{duarte2011and} study children's interactions at different stages of the information seeking process.

The literature that we have discussed has been pivotal to the advancement of IR regarding children and search engines. Yet, research has been sporadic throughout the last two decades and limited to few research groups. We attribute the inconsistent nature of this type of research to issues related to obtaining children's data and the ability to freely share benchmarks as well as labeled data to inform development and evaluation, which other domains do not have with as datasets are fallible, e.g., CLEF, TREC, etc. More importantly, these scholarly works focus on children in specific age ranges, e.g., 8 to 11 year-olds or adolescents. This is a concern as (i) children in these user groups do not exhibit uniform (search) behavior, and (ii) the full spectrum of requirements and expectations of children (18 and under) has been overlooked. Given the exigency of supporting children's search, a one-size-fits-all solution is not adequate. Instead, various lenses that take into consideration the different skills children possess should be pursued when altering or designing search systems. In this paper, we spotlight the \textit{accessibility} lens. Explicitly, we discuss the demand for making search engines more \textit{accessible} for children of different mental, physical, and learning/developmental abilities.

\section{"We got to get this wagon train a-movin'!": Gaps}

The call to create accessible search engines is not a new idea \cite{serdyukov2011towards, kerkmann2012accessibility}. IR researchers have allocated efforts to explore how adult searchers affected by dyslexia deal with search engines \cite{fourney2018assessing}, the information-seeking process of adults with invisible disabilities \cite{muir2019considering}, and how to best adapt search engines to searchers who are vision-impaired \cite{vtyurina2019towards,vtyurina2019verse}. Unfortunately, substantial progress has yet to be made by the IR community regarding accessibility for children--the population currently under study--beyond some preliminary investigations like autism \cite{young2016children}. To highlight some opportunities for adapting search engines to support children with disabilities, we provide some motivating examples showcasing the necessity to dedicate scholarly efforts to holistically look at children, search engines, and accessibility.

\textbf{Query formulation}. \citet{downs2020kidspell} present a specialized spellchecker (KidSpell) tailored explicitly to children to help with query formulation, as there are trends in how children misspell words that differ from those observed among adults. We posit, however, that KidSpell would not necessarily work for children with dyslexia, who are prone to make different types of misspellings \cite{li2013polispell,kvikne2019search}. Additionally, KidSpell incorporates a visual and audio component that may adversely impact users with autism by unintentionally inducing sensory overload \cite{mikropoulos2020acceptance}.
 
\textbf{Query suggestions.} \citet{madrazo2018looking} introduce ReQuIK, a query suggestion strategy for children ages 8 to 11 that infers children's intent from their original queries. ReQuIK, which conventionally displays suggestions, may not be applicable for children who are vision impaired. Vision hurdles could indeed be bypassed by relying on conversational voice search, which \citet{lovato2019hey} explore, but this too comes with accessibility gaps. For a conversational voice search to be effective, it must understand the verbal input provided by users, which is difficult in the case of children in general \cite{gossen2016search,gossen2013voice,shivakumar2020transfer,fainberg2016improving,yarosh2018children,cheng2018doesn}, but is further exacerbated when children have speech impediments or disorders like Tourette syndrome. 
 
\textbf{Ranking.} \citet{milton2020korsce} and \citet{usta2021learning} propose new ranking algorithms to prioritize retrieved results that are relevant to educational contexts. Both rankers depend upon a user's age, which is mapped to a respective school grade and then used to identify online resources that align with the educational curriculum outlined for a said grade. By assuming that a child's age directly matches with their developmental stage or reading ability, the rankers cannot account for children that have disabilities like autism or dyslexia.

\textbf{Interactions.} \citet{landoni2021right} investigate whether adding visual cues to SERP displays to provide clues for resource relevance eases the process of locating resources addressing children's information needs. Aside from the visual issues that we have already discussed, the authors utilize a traditional SERP which can be difficult to navigate for young children who lack fine motor skills \cite{gossen2016search}. When considering children with more extreme physical disabilities, like missing limbs, cerebral palsy, or muscular dystrophy, it becomes almost impossible to employ standard keyboards and mouses to interact with SERP. 

\textbf{SERP.} \citet{landoni2020inside} explore the emotional responses captured in SERP generated by children searching for classroom-related materials. The authors point out that during their user study, teachers observed emotional responses evoked in child participants by search results. The ethical nature of whether distressing search tasks are suitable for children on an emotional level is also brought up. This concern is escalated when considering children with mental health disorders, particularly those with bipolar, depressive, anxiety, attention deficit (ADD/ADHD), or post-traumatic stress disorders. Additionally, research strategies that leverage the emotions of children must also consider that some of them may not have an adequate emotional maturity level or ability to feel emotions, which is a symptom of antisocial personality disorder, psychopathy, or sociopathy.

The examples we have discussed neither in any way represent an exhaustive list of all known disabilities that impact children nor the open problems related to the accessibility of search engines. Instead, they are meant to offer a glimpse of how existing research is underserving children with disabilities and their potential use of search engines.

\section{"Well, then, let's find out together!": Future}

Even from the few examples we have drawn attention to in Section 2, it is clear that there are plenty of research gaps that emerge when considering the complex problem of accessibility. Indeed, living up to the requirements and expectations of how children with disabilities use search engines and how search engines can better support them, is not an easy feat. We hypothesize that research in this area cannot be done in isolation and that instead, it will require long-term, dedicated collaborations involving researchers from multiple disciplines. Increasing accessibility of search engines is one important step forward towards creating an environment where all users, especially children, have equal access to essential information. Accessibility is a broad concept, one that encompasses multiple lenses. As a starting point to foster discussion in this area, we shine a light on three categories of disabilities that affect children's ability to use search engines: physical, mental, and developmental/learning. 

Consider, for instance, physical interfaces, e.g., augmentative communication devices or voice synthesizers, that are meant to convey ideas for disabled people -- how do researchers ensure users can use such devices to interface with search engines as a means for users to form queries? How do search engines handle input received from any assistive hardware and translate users' information needs into queries or trigger click interactions? While some hardware has already been designed, e.g., specialized controllers in the video game industry, to improve accessibility for users with motor disabilities, what can we learn as a community from these specialized tools and adopt to make search more accessible? What about the way that SERP are presented, would spacing of results be an issue? Do audio and visual aspects of SERP present barriers or distractions for children with disorders such as anxiety or ADD/ADHD?

Many interfaces on child-oriented search engines are designed to be more engaging by using audio and visual elements. Some of these visual elements are based on bright colors, which are known to have inherent emotions attached to them, e.g., red is associated with sweets and green with healthy \cite{huang2015eat}. What sort of impact do these visual elements have on children with mental disorders? Black and other dark colors are often associated with connotations of depression. How do search engines deal with depressed children as users? How would a search engine know if a child is suffering from some form of trauma that could be triggered by results presented to them? Many mental health issues will go undiagnosed for years. How can a search engine be aware of and adapt to something that even the users, their teachers, or their parents do not know?

With developmental conditions where age and maturity differ, how do search engines account for the difference in development? In their current state, modern search engines are heavily text-based, requiring that users can adequately read and understand the resources returned. When a child is behind in their reading level or math skills due to dyslexia or dyscalculia, how do search engines adapt so that they can provide resources that are at children's cognitive level instead of age level? 

The questions we have expounded here are not comprehensive of all the avenues to explore. Instead, our goal is simply to encourage researchers to consider significant barriers that, nowadays, search engines exhibit when it comes to serving children with disabilities. As previously mentioned, tackling such a complex and diverse matter should take place in a multidisciplinary setting. Together, researchers from domains including education, mechanical \& electrical engineering, developmental \& clinical psychology, sociology, natural language processing, human-computer interaction, machine learning, and medical must ruminate on what accessibility means to them. It is only based on their reflection and collaboration with IR researchers that will enable the design, development, and assessment of search engines that respond to real-world necessities and can therefore provide the appropriate support that children with disabilities need when interacting with search engines. It is key to remember that accessibility does not have to be thought of as a burden, as it enhances search engines without reducing their effectiveness \cite{azzopardi2010relationship}. While the current generation of IR researchers has focused on the effectiveness of search engines, it is imperative to inspire the next to focus on accessibility as it is a long-term problem. It is also imperative that we involve researchers coming into the IR community to (i) successfully be part of interdisciplinary research teams and (ii) update IR curriculum taught at undergraduate and graduate institutions to increase awareness about different types of users and accessibility issues affecting IR-related systems --beyond search engines--as they will be the ones carrying the torch forward. As Eve Andersson, the Director of Google said, ``the accessibility problems of today are the mainstream breakthroughs of tomorrow'' \cite{brownlee_2018}.

\begin{acks}
Work partially funded by NSF Awards \# 1763649 and \#1930464. 
\end{acks}

\bibliographystyle{ACM-Reference-Format}
\bibliography{VisionPaper}


\begin{thebibliography}{54}


\ifx \showCODEN    \undefined \def \showCODEN     #1{\unskip}     \fi
\ifx \showDOI      \undefined \def \showDOI       #1{#1}\fi
\ifx \showISBNx    \undefined \def \showISBNx     #1{\unskip}     \fi
\ifx \showISBNxiii \undefined \def \showISBNxiii  #1{\unskip}     \fi
\ifx \showISSN     \undefined \def \showISSN      #1{\unskip}     \fi
\ifx \showLCCN     \undefined \def \showLCCN      #1{\unskip}     \fi
\ifx \shownote     \undefined \def \shownote      #1{#1}          \fi
\ifx \showarticletitle \undefined \def \showarticletitle #1{#1}   \fi
\ifx \showURL      \undefined \def \showURL       {\relax}        \fi
\providecommand\bibfield[2]{#2}
\providecommand\bibinfo[2]{#2}
\providecommand\natexlab[1]{#1}
\providecommand\showeprint[2][]{arXiv:#2}

\bibitem[\protect\citeauthoryear{??}{pew}{2021}]%
        {pewSE}
 \bibinfo{year}{2021}\natexlab{}.
\newblock \bibinfo{title}{Demographics of Internet and Home Broadband Usage in
  the United States}.
\newblock
\newblock
\urldef\tempurl%
\url{https://www.pewresearch.org/internet/fact-sheet/internet-broadband/?menuItem=6b886b10-55ec-44bc-b5a4-740f5366a404}
\showURL{%
\tempurl}


\bibitem[\protect\citeauthoryear{Anuyah, Milton, Green, and Pera}{Anuyah
  et~al\mbox{.}}{2019}]%
        {anuyah2019classroom}
\bibfield{author}{\bibinfo{person}{Oghenemaro Anuyah}, \bibinfo{person}{Ashlee
  Milton}, \bibinfo{person}{Michael Green}, {and}
  \bibinfo{person}{Maria~Soledad Pera}.} \bibinfo{year}{2019}\natexlab{}.
\newblock \showarticletitle{An empirical analysis of search engines’ response
  to web search queries associated with the classroom setting}.
\newblock \bibinfo{journal}{\emph{Aslib Journal of Information Management}}
  (\bibinfo{year}{2019}).
\newblock


\bibitem[\protect\citeauthoryear{Azzopardi and Bache}{Azzopardi and
  Bache}{2010}]%
        {azzopardi2010relationship}
\bibfield{author}{\bibinfo{person}{Leif Azzopardi} {and}
  \bibinfo{person}{Richard Bache}.} \bibinfo{year}{2010}\natexlab{}.
\newblock \showarticletitle{On the relationship between effectiveness and
  accessibility}. In \bibinfo{booktitle}{\emph{Proceedings of the 33rd
  international ACM SIGIR conference on Research and development in information
  retrieval}}. \bibinfo{pages}{889--890}.
\newblock


\bibitem[\protect\citeauthoryear{Bilal}{Bilal}{1998}]%
        {bilal1998children}
\bibfield{author}{\bibinfo{person}{Dania Bilal}.}
  \bibinfo{year}{1998}\natexlab{}.
\newblock \showarticletitle{Children's Search Processes in Using World Wide Web
  Search Engines: An Exploratory Study.}. In
  \bibinfo{booktitle}{\emph{Proceedings of the ASIS annual meeting}},
  Vol.~\bibinfo{volume}{35}. ERIC, \bibinfo{pages}{45--53}.
\newblock


\bibitem[\protect\citeauthoryear{Bilal}{Bilal}{1999}]%
        {bilal1999web}
\bibfield{author}{\bibinfo{person}{Dania Bilal}.}
  \bibinfo{year}{1999}\natexlab{}.
\newblock \showarticletitle{Web Search Engines for Children: A Comparative
  Study and Performance Evaluation of" Yahooligans!,"" Ask Jeeves for Kids,"
  and" Super Snooper.".}. In \bibinfo{booktitle}{\emph{Proceedings of the ASIS
  annual meeting}}, Vol.~\bibinfo{volume}{36}. ERIC, \bibinfo{pages}{70--83}.
\newblock


\bibitem[\protect\citeauthoryear{Bilal}{Bilal}{2010}]%
        {bilal2010mediated}
\bibfield{author}{\bibinfo{person}{Dania Bilal}.}
  \bibinfo{year}{2010}\natexlab{}.
\newblock \showarticletitle{The mediated information needs of children on the
  Autism Spectrum Disorder (ASD)}. In \bibinfo{booktitle}{\emph{Proceedings of
  the 31st ACM SIGIR Workshop on Accessible Search Systems, Geneva,
  Switzerland}}. Citeseer, \bibinfo{pages}{42--49}.
\newblock


\bibitem[\protect\citeauthoryear{Bilal and Boehm}{Bilal and Boehm}{2017}]%
        {bilal2017towards}
\bibfield{author}{\bibinfo{person}{Dania Bilal} {and} \bibinfo{person}{Meredith
  Boehm}.} \bibinfo{year}{2017}\natexlab{}.
\newblock \showarticletitle{Towards new methodologies for assessing relevance
  of information retrieval from web search engines on children’s queries}.
\newblock \bibinfo{journal}{\emph{Qualitative and Quantitative Methods in
  Libraries}} \bibinfo{volume}{2}, \bibinfo{number}{1} (\bibinfo{year}{2017}),
  \bibinfo{pages}{93--100}.
\newblock


\bibitem[\protect\citeauthoryear{Bilal and Gwizdka}{Bilal and Gwizdka}{2018}]%
        {bilal2018children}
\bibfield{author}{\bibinfo{person}{Dania Bilal} {and} \bibinfo{person}{Jacek
  Gwizdka}.} \bibinfo{year}{2018}\natexlab{}.
\newblock \showarticletitle{Children's query types and reformulations in Google
  search}.
\newblock \bibinfo{journal}{\emph{Information Processing \& Management}}
  \bibinfo{volume}{54}, \bibinfo{number}{6} (\bibinfo{year}{2018}),
  \bibinfo{pages}{1022--1041}.
\newblock


\bibitem[\protect\citeauthoryear{Bilal and Huang}{Bilal and Huang}{2019}]%
        {bilal2019readability}
\bibfield{author}{\bibinfo{person}{Dania Bilal} {and} \bibinfo{person}{Li-Min
  Huang}.} \bibinfo{year}{2019}\natexlab{}.
\newblock \showarticletitle{Readability and word complexity of SERPs snippets
  and web pages on children’s search queries}.
\newblock \bibinfo{journal}{\emph{Aslib Journal of Information Management}}
  (\bibinfo{year}{2019}).
\newblock


\bibitem[\protect\citeauthoryear{Brownlee}{Brownlee}{2018}]%
        {brownlee_2018}
\bibfield{author}{\bibinfo{person}{John Brownlee}.}
  \bibinfo{year}{2018}\natexlab{}.
\newblock \bibinfo{title}{How Designing For Disabled People Is Giving Google An
  Edge}.
\newblock
\newblock
\urldef\tempurl%
\url{https://www.fastcompany.com/3060090/how-designing-for-the-disabled-is-giving-google-an-edge}
\showURL{%
\tempurl}


\bibitem[\protect\citeauthoryear{Chaffey}{Chaffey}{2021}]%
        {SEstats}
\bibfield{author}{\bibinfo{person}{Dave Chaffey}.}
  \bibinfo{year}{2021}\natexlab{}.
\newblock \bibinfo{title}{Search engine marketing statistics: The latest search
  usage and adoption data to inform your strategies and tactics}.
\newblock
\newblock
\urldef\tempurl%
\url{https://www.smartinsights.com/search-engine-marketing/search-engine-statistics/}
\showURL{%
\tempurl}


\bibitem[\protect\citeauthoryear{Cheng, Yen, Chen, Chen, and Hiniker}{Cheng
  et~al\mbox{.}}{2018}]%
        {cheng2018doesn}
\bibfield{author}{\bibinfo{person}{Yi Cheng}, \bibinfo{person}{Kate Yen},
  \bibinfo{person}{Yeqi Chen}, \bibinfo{person}{Sijin Chen}, {and}
  \bibinfo{person}{Alexis Hiniker}.} \bibinfo{year}{2018}\natexlab{}.
\newblock \showarticletitle{Why doesn't it work? voice-driven interfaces and
  young children's communication repair strategies}. In
  \bibinfo{booktitle}{\emph{Proceedings of the 17th ACM Conference on
  Interaction Design and Children}}. \bibinfo{pages}{337--348}.
\newblock


\bibitem[\protect\citeauthoryear{Collins-Thompson, Bennett, White, De~La~Chica,
  and Sontag}{Collins-Thompson et~al\mbox{.}}{2011}]%
        {collins2011personalizing}
\bibfield{author}{\bibinfo{person}{Kevyn Collins-Thompson},
  \bibinfo{person}{Paul~N Bennett}, \bibinfo{person}{Ryen~W White},
  \bibinfo{person}{Sebastian De~La~Chica}, {and} \bibinfo{person}{David
  Sontag}.} \bibinfo{year}{2011}\natexlab{}.
\newblock \showarticletitle{Personalizing web search results by reading level}.
  In \bibinfo{booktitle}{\emph{Proceedings of the 20th ACM international
  conference on Information and knowledge management}}.
  \bibinfo{pages}{403--412}.
\newblock


\bibitem[\protect\citeauthoryear{Danovitch}{Danovitch}{2019}]%
        {danovitch2019growing}
\bibfield{author}{\bibinfo{person}{Judith~H Danovitch}.}
  \bibinfo{year}{2019}\natexlab{}.
\newblock \showarticletitle{Growing up with Google: How children's
  understanding and use of internet-based devices relates to cognitive
  development}.
\newblock \bibinfo{journal}{\emph{Human Behavior and Emerging Technologies}}
  \bibinfo{volume}{1}, \bibinfo{number}{2} (\bibinfo{year}{2019}),
  \bibinfo{pages}{81--90}.
\newblock


\bibitem[\protect\citeauthoryear{Downs, Anuyah, Shukla, Fails, Pera, Wright,
  and Kennington}{Downs et~al\mbox{.}}{2020}]%
        {downs2020kidspell}
\bibfield{author}{\bibinfo{person}{Brody Downs}, \bibinfo{person}{Oghenemaro
  Anuyah}, \bibinfo{person}{Aprajita Shukla}, \bibinfo{person}{Jerry~Alan
  Fails}, \bibinfo{person}{Sole Pera}, \bibinfo{person}{Katherine Wright},
  {and} \bibinfo{person}{Casey Kennington}.} \bibinfo{year}{2020}\natexlab{}.
\newblock \showarticletitle{Kidspell: A child-oriented, rule-based, phonetic
  spellchecker}. In \bibinfo{booktitle}{\emph{Proceedings of The 12th Language
  Resources and Evaluation Conference}}. \bibinfo{pages}{6937--6946}.
\newblock


\bibitem[\protect\citeauthoryear{Dragovic, Madrazo~Azpiazu, and Pera}{Dragovic
  et~al\mbox{.}}{2016}]%
        {dragovic2016sven}
\bibfield{author}{\bibinfo{person}{Nevena Dragovic}, \bibinfo{person}{Ion
  Madrazo~Azpiazu}, {and} \bibinfo{person}{Maria~Soledad Pera}.}
  \bibinfo{year}{2016}\natexlab{}.
\newblock \showarticletitle{" Is Sven Seven?" A Search Intent Module for
  Children}. In \bibinfo{booktitle}{\emph{Proceedings of the 39th International
  ACM SIGIR conference on Research and Development in Information Retrieval}}.
  \bibinfo{pages}{885--888}.
\newblock


\bibitem[\protect\citeauthoryear{Duarte~Torres and Weber}{Duarte~Torres and
  Weber}{2011}]%
        {duarte2011and}
\bibfield{author}{\bibinfo{person}{Sergio Duarte~Torres} {and}
  \bibinfo{person}{Ingmar Weber}.} \bibinfo{year}{2011}\natexlab{}.
\newblock \showarticletitle{What and how children search on the web}. In
  \bibinfo{booktitle}{\emph{Proceedings of the 20th ACM international
  conference on Information and knowledge management}}.
  \bibinfo{pages}{393--402}.
\newblock


\bibitem[\protect\citeauthoryear{Eickhoff, Serdyukov, and de~Vries}{Eickhoff
  et~al\mbox{.}}{2010}]%
        {eickhoff2010suitability}
\bibfield{author}{\bibinfo{person}{Carsten Eickhoff}, \bibinfo{person}{Pavel
  Serdyukov}, {and} \bibinfo{person}{Arjen~P. de Vries}.}
  \bibinfo{year}{2010}\natexlab{}.
\newblock \showarticletitle{Web Page Classification on Child Suitability}. In
  \bibinfo{booktitle}{\emph{Proceedings of the 19th ACM International
  Conference on Information and Knowledge Management}} (Toronto, ON, Canada)
  \emph{(\bibinfo{series}{CIKM ’10})}. \bibinfo{publisher}{Association for
  Computing Machinery}, \bibinfo{address}{New York, NY, USA},
  \bibinfo{pages}{1425–1428}.
\newblock
\showISBNx{9781450300995}
\urldef\tempurl%
\url{https://doi.org/10.1145/1871437.1871638}
\showDOI{\tempurl}


\bibitem[\protect\citeauthoryear{Fainberg, Bell, Lincoln, and Renals}{Fainberg
  et~al\mbox{.}}{2016}]%
        {fainberg2016improving}
\bibfield{author}{\bibinfo{person}{Joachim Fainberg}, \bibinfo{person}{Peter
  Bell}, \bibinfo{person}{Mike Lincoln}, {and} \bibinfo{person}{Steve Renals}.}
  \bibinfo{year}{2016}\natexlab{}.
\newblock \showarticletitle{Improving Children's Speech Recognition Through
  Out-of-Domain Data Augmentation.}. In
  \bibinfo{booktitle}{\emph{Interspeech}}. \bibinfo{pages}{1598--1602}.
\newblock


\bibitem[\protect\citeauthoryear{Figueiredo and Meyers}{Figueiredo and
  Meyers}{2019}]%
        {figueiredo2019false}
\bibfield{author}{\bibinfo{person}{Vanessa Figueiredo} {and}
  \bibinfo{person}{Eric~M Meyers}.} \bibinfo{year}{2019}\natexlab{}.
\newblock \showarticletitle{The false trade-off of relevance for safety in
  children's search systems}.
\newblock \bibinfo{journal}{\emph{Proceedings of the Association for
  Information Science and Technology}} \bibinfo{volume}{56},
  \bibinfo{number}{1} (\bibinfo{year}{2019}), \bibinfo{pages}{651--653}.
\newblock


\bibitem[\protect\citeauthoryear{Foss, Druin, Brewer, Lo, Sanchez, Golub, and
  Hutchinson}{Foss et~al\mbox{.}}{2012}]%
        {foss2012children}
\bibfield{author}{\bibinfo{person}{Elizabeth Foss}, \bibinfo{person}{Allison
  Druin}, \bibinfo{person}{Robin Brewer}, \bibinfo{person}{Phillip Lo},
  \bibinfo{person}{Luis Sanchez}, \bibinfo{person}{Evan Golub}, {and}
  \bibinfo{person}{Hilary Hutchinson}.} \bibinfo{year}{2012}\natexlab{}.
\newblock \showarticletitle{Children's search roles at home: Implications for
  designers, researchers, educators, and parents}.
\newblock \bibinfo{journal}{\emph{Journal of the American Society for
  Information Science and Technology}} \bibinfo{volume}{63},
  \bibinfo{number}{3} (\bibinfo{year}{2012}), \bibinfo{pages}{558--573}.
\newblock


\bibitem[\protect\citeauthoryear{Fourney, Ringel~Morris, Ali, and
  Vonessen}{Fourney et~al\mbox{.}}{2018}]%
        {fourney2018assessing}
\bibfield{author}{\bibinfo{person}{Adam Fourney}, \bibinfo{person}{Meredith
  Ringel~Morris}, \bibinfo{person}{Abdullah Ali}, {and} \bibinfo{person}{Laura
  Vonessen}.} \bibinfo{year}{2018}\natexlab{}.
\newblock \showarticletitle{Assessing the readability of web search results for
  searchers with dyslexia}. In \bibinfo{booktitle}{\emph{The 41st International
  ACM SIGIR Conference on Research \& Development in Information Retrieval}}.
  \bibinfo{pages}{1069--1072}.
\newblock


\bibitem[\protect\citeauthoryear{Gossen}{Gossen}{2016}]%
        {gossen2016search}
\bibfield{author}{\bibinfo{person}{Tatiana Gossen}.}
  \bibinfo{year}{2016}\natexlab{}.
\newblock \bibinfo{booktitle}{\emph{Search engines for children: search user
  interfaces and information-seeking behaviour}}.
\newblock \bibinfo{publisher}{Springer}.
\newblock


\bibitem[\protect\citeauthoryear{Gossen, Kotzyba, Stober, and
  N{\"u}rnberger}{Gossen et~al\mbox{.}}{2013}]%
        {gossen2013voice}
\bibfield{author}{\bibinfo{person}{Tatiana Gossen}, \bibinfo{person}{Michael
  Kotzyba}, \bibinfo{person}{Sebastian Stober}, {and} \bibinfo{person}{Andreas
  N{\"u}rnberger}.} \bibinfo{year}{2013}\natexlab{}.
\newblock \showarticletitle{Voice-controlled search user interfaces for young
  users}. In \bibinfo{booktitle}{\emph{7th Annual Symposium on Human-Computer
  Interaction and Information Retrieval}}. \bibinfo{pages}{2--5}.
\newblock


\bibitem[\protect\citeauthoryear{Gupta and Hilal}{Gupta and Hilal}{2013}]%
        {gupta2013algorithm}
\bibfield{author}{\bibinfo{person}{Neha Gupta} {and} \bibinfo{person}{Saba
  Hilal}.} \bibinfo{year}{2013}\natexlab{}.
\newblock \showarticletitle{Algorithm to filter \& redirect the web content for
  kids’}.
\newblock \bibinfo{journal}{\emph{International Journal of Engineering and
  Technology}}  \bibinfo{volume}{5} (\bibinfo{year}{2013}).
\newblock


\bibitem[\protect\citeauthoryear{Gwizdka and Bilal}{Gwizdka and Bilal}{2017}]%
        {gwizdka2017analysis}
\bibfield{author}{\bibinfo{person}{Jacek Gwizdka} {and} \bibinfo{person}{Dania
  Bilal}.} \bibinfo{year}{2017}\natexlab{}.
\newblock \showarticletitle{Analysis of children's queries and click behavior
  on ranked results and their thought processes in google search}. In
  \bibinfo{booktitle}{\emph{Proceedings of the 2017 conference on conference
  human information interaction and retrieval}}.
  \bibinfo{publisher}{Association for Computing Machinery},
  \bibinfo{address}{New York, NY, USA}, \bibinfo{pages}{377--380}.
\newblock


\bibitem[\protect\citeauthoryear{Haider and Sundin}{Haider and Sundin}{2019}]%
        {haider2019invisible}
\bibfield{author}{\bibinfo{person}{Jutta Haider} {and} \bibinfo{person}{Olof
  Sundin}.} \bibinfo{year}{2019}\natexlab{}.
\newblock \bibinfo{booktitle}{\emph{Invisible search and online search engines:
  The ubiquity of search in everyday life}}.
\newblock \bibinfo{publisher}{Routledge}.
\newblock


\bibitem[\protect\citeauthoryear{Huang and Lu}{Huang and Lu}{2015}]%
        {huang2015eat}
\bibfield{author}{\bibinfo{person}{L Huang} {and} \bibinfo{person}{J Lu}.}
  \bibinfo{year}{2015}\natexlab{}.
\newblock \showarticletitle{Eat with your eyes: Package color influences the
  expectation of food taste and healthiness moderated by external eating}.
\newblock \bibinfo{journal}{\emph{Marketing Management}} \bibinfo{volume}{25},
  \bibinfo{number}{2} (\bibinfo{year}{2015}), \bibinfo{pages}{71--87}.
\newblock


\bibitem[\protect\citeauthoryear{Kerkmann and Lewandowski}{Kerkmann and
  Lewandowski}{2012}]%
        {kerkmann2012accessibility}
\bibfield{author}{\bibinfo{person}{Friederike Kerkmann} {and}
  \bibinfo{person}{Dirk Lewandowski}.} \bibinfo{year}{2012}\natexlab{}.
\newblock \showarticletitle{Accessibility of web search engines: Towards a
  deeper understanding of barriers for people with disabilities}.
\newblock \bibinfo{journal}{\emph{Library review}} (\bibinfo{year}{2012}).
\newblock


\bibitem[\protect\citeauthoryear{Kvikne and Berget}{Kvikne and Berget}{2019}]%
        {kvikne2019search}
\bibfield{author}{\bibinfo{person}{Birgit Kvikne} {and} \bibinfo{person}{Gerd
  Berget}.} \bibinfo{year}{2019}\natexlab{}.
\newblock \showarticletitle{In search of trustworthy information: a qualitative
  study of the search behavior of people with dyslexia in Norway}.
\newblock \bibinfo{journal}{\emph{Universal Access in the Information Society}}
  (\bibinfo{year}{2019}), \bibinfo{pages}{1--12}.
\newblock


\bibitem[\protect\citeauthoryear{Landoni, Aliannejadi, Huibers, Murgia, and
  Pera}{Landoni et~al\mbox{.}}{2021}]%
        {landoni2021right}
\bibfield{author}{\bibinfo{person}{Monica Landoni}, \bibinfo{person}{Mohammad
  Aliannejadi}, \bibinfo{person}{Theo Huibers}, \bibinfo{person}{Emiliana
  Murgia}, {and} \bibinfo{person}{Maria~Soledad Pera}.}
  \bibinfo{year}{2021}\natexlab{}.
\newblock \showarticletitle{Right Way, Right Time: Towards a Better
  Comprehension of Young Students' Needs When Looking for Relevant Search
  Results}. In \bibinfo{booktitle}{\emph{Proceedings of 29th Conference on User
  Modeling, Adaptation and Personalization (UMAP)}}. \bibinfo{pages}{To
  Appear}.
\newblock


\bibitem[\protect\citeauthoryear{Landoni, Matteri, Murgia, Huibers, and
  Pera}{Landoni et~al\mbox{.}}{2019}]%
        {landoni2019sonny}
\bibfield{author}{\bibinfo{person}{Monica Landoni}, \bibinfo{person}{Davide
  Matteri}, \bibinfo{person}{Emiliana Murgia}, \bibinfo{person}{Theo Huibers},
  {and} \bibinfo{person}{Maria~Soledad Pera}.} \bibinfo{year}{2019}\natexlab{}.
\newblock \showarticletitle{Sonny, Cerca! evaluating the impact of using a
  vocal assistant to search at school}. In
  \bibinfo{booktitle}{\emph{International Conference of the Cross-Language
  Evaluation Forum for European Languages}}. Springer,
  \bibinfo{publisher}{"Springer International Publishing"},
  \bibinfo{address}{"Cham"}, \bibinfo{pages}{101--113}.
\newblock


\bibitem[\protect\citeauthoryear{Landoni, Pera, Murgia, and Huibers}{Landoni
  et~al\mbox{.}}{2020}]%
        {landoni2020inside}
\bibfield{author}{\bibinfo{person}{Monica Landoni},
  \bibinfo{person}{Maria~Soledad Pera}, \bibinfo{person}{Emiliana Murgia},
  {and} \bibinfo{person}{Theo Huibers}.} \bibinfo{year}{2020}\natexlab{}.
\newblock \showarticletitle{Inside Out: Exploring the Emotional Side of Search
  Engines in the Classroom}. In \bibinfo{booktitle}{\emph{Proceedings of the
  28th ACM Conference on User Modeling, Adaptation and Personalization}}.
  \bibinfo{pages}{136--144}.
\newblock


\bibitem[\protect\citeauthoryear{Li, Sbattella, and Tedesco}{Li
  et~al\mbox{.}}{2013}]%
        {li2013polispell}
\bibfield{author}{\bibinfo{person}{Alberto~Quattrini Li},
  \bibinfo{person}{Licia Sbattella}, {and} \bibinfo{person}{Roberto Tedesco}.}
  \bibinfo{year}{2013}\natexlab{}.
\newblock \showarticletitle{Polispell: an adaptive spellchecker and predictor
  for people with dyslexia}. In \bibinfo{booktitle}{\emph{International
  Conference on User Modeling, Adaptation, and Personalization}}. Springer,
  \bibinfo{pages}{302--309}.
\newblock


\bibitem[\protect\citeauthoryear{Lovato, Piper, and Wartella}{Lovato
  et~al\mbox{.}}{2019}]%
        {lovato2019hey}
\bibfield{author}{\bibinfo{person}{Silvia~B Lovato},
  \bibinfo{person}{Anne~Marie Piper}, {and} \bibinfo{person}{Ellen~A
  Wartella}.} \bibinfo{year}{2019}\natexlab{}.
\newblock \showarticletitle{Hey Google, do unicorns exist? Conversational
  agents as a path to answers to children's questions}. In
  \bibinfo{booktitle}{\emph{Proceedings of the 18th ACM International
  Conference on Interaction Design and Children}}. \bibinfo{pages}{301--313}.
\newblock


\bibitem[\protect\citeauthoryear{Madrazo~Azpiazu, Dragovic, Anuyah, and
  Pera}{Madrazo~Azpiazu et~al\mbox{.}}{2018}]%
        {madrazo2018looking}
\bibfield{author}{\bibinfo{person}{Ion Madrazo~Azpiazu},
  \bibinfo{person}{Nevena Dragovic}, \bibinfo{person}{Oghenemaro Anuyah}, {and}
  \bibinfo{person}{Maria~Soledad Pera}.} \bibinfo{year}{2018}\natexlab{}.
\newblock \showarticletitle{Looking for the Movie Seven or Sven from the Movie
  Frozen? A Multi-perspective Strategy for Recommending Queries for Children}.
  In \bibinfo{booktitle}{\emph{Proceedings of the 2018 Conference on Human
  Information Interaction \& Retrieval}}. \bibinfo{pages}{92--101}.
\newblock


\bibitem[\protect\citeauthoryear{Mikropoulos, Delimitros, Gaintatzis, Iatraki,
  Stergiouli, Tsiara, and Kalyvioti}{Mikropoulos et~al\mbox{.}}{2020}]%
        {mikropoulos2020acceptance}
\bibfield{author}{\bibinfo{person}{Tassos~A Mikropoulos},
  \bibinfo{person}{Michael Delimitros}, \bibinfo{person}{Pavlos Gaintatzis},
  \bibinfo{person}{Georgia Iatraki}, \bibinfo{person}{Aikaterini Stergiouli},
  \bibinfo{person}{Angeliki Tsiara}, {and} \bibinfo{person}{Katerina
  Kalyvioti}.} \bibinfo{year}{2020}\natexlab{}.
\newblock \showarticletitle{Acceptance and User Experience of an Augmented
  Reality System for the Simulation of Sensory Overload in Children with
  Autism}. In \bibinfo{booktitle}{\emph{2020 6th International Conference of
  the Immersive Learning Research Network (iLRN)}}. IEEE,
  \bibinfo{pages}{86--92}.
\newblock


\bibitem[\protect\citeauthoryear{Milton, Anuyah, Spear, Wright, and
  Pera}{Milton et~al\mbox{.}}{2020}]%
        {milton2020korsce}
\bibfield{author}{\bibinfo{person}{Ashlee Milton}, \bibinfo{person}{Oghenemaro
  Anuyah}, \bibinfo{person}{Lawrence Spear}, \bibinfo{person}{Katherine~Landau
  Wright}, {and} \bibinfo{person}{Maria~Soledad Pera}.}
  \bibinfo{year}{2020}\natexlab{}.
\newblock \showarticletitle{A Ranking Strategy to Promote Resources Supporting
  the Classroom Environment}. In \bibinfo{booktitle}{\emph{To appear in the
  Proceedings of the 2020 IEEE/WIC/ACM International Joint Conference on Web
  Intelligence and Intelligent Agent Technology (WI-IAT'20)}}.
\newblock


\bibitem[\protect\citeauthoryear{Muir, Thompson, and Qayyum}{Muir
  et~al\mbox{.}}{2019}]%
        {muir2019considering}
\bibfield{author}{\bibinfo{person}{Rebecca Muir}, \bibinfo{person}{Kim~M
  Thompson}, {and} \bibinfo{person}{Asim Qayyum}.}
  \bibinfo{year}{2019}\natexlab{}.
\newblock \showarticletitle{Considering “atmosphere” in facilitating
  information seeking by people with invisible disabilities in public
  libraries}.
\newblock \bibinfo{journal}{\emph{Proceedings of the Association for
  Information Science and Technology}} \bibinfo{volume}{56},
  \bibinfo{number}{1} (\bibinfo{year}{2019}), \bibinfo{pages}{216--226}.
\newblock


\bibitem[\protect\citeauthoryear{Patel and Singh}{Patel and Singh}{2016}]%
        {patel2016kids}
\bibfield{author}{\bibinfo{person}{Deepshikha Patel} {and}
  \bibinfo{person}{Prashant~Kumar Singh}.} \bibinfo{year}{2016}\natexlab{}.
\newblock \showarticletitle{Kids safe search classification model}. In
  \bibinfo{booktitle}{\emph{2016 International Conference on Communication and
  Electronics Systems (ICCES)}}. IEEE, \bibinfo{pages}{1--7}.
\newblock


\bibitem[\protect\citeauthoryear{Rowlands, Nicholas, Williams, Huntington,
  Fieldhouse, Gunter, Withey, Jamali, Dobrowolski, and Tenopir}{Rowlands
  et~al\mbox{.}}{2008}]%
        {rowlands2008google}
\bibfield{author}{\bibinfo{person}{Ian Rowlands}, \bibinfo{person}{David
  Nicholas}, \bibinfo{person}{Peter Williams}, \bibinfo{person}{Paul
  Huntington}, \bibinfo{person}{Maggie Fieldhouse}, \bibinfo{person}{Barrie
  Gunter}, \bibinfo{person}{Richard Withey}, \bibinfo{person}{Hamid~R Jamali},
  \bibinfo{person}{Tom Dobrowolski}, {and} \bibinfo{person}{Carol Tenopir}.}
  \bibinfo{year}{2008}\natexlab{}.
\newblock \showarticletitle{The Google generation: the information behaviour of
  the researcher of the future}. In \bibinfo{booktitle}{\emph{Aslib
  proceedings}}. Emerald Group Publishing Limited.
\newblock


\bibitem[\protect\citeauthoryear{Sentance}{Sentance}{[n.d.]}]%
        {sentance}
\bibfield{author}{\bibinfo{person}{Rebecca Sentance}.}
  \bibinfo{year}{[n.d.]}\natexlab{}.
\newblock \bibinfo{title}{What are the differences in how age demographics
  search the internet?}
\newblock
\newblock
\urldef\tempurl%
\url{https://www.userzoom.com/ux-library/what-are-the-differences-in-how-age-demographics-search/}
\showURL{%
\tempurl}


\bibitem[\protect\citeauthoryear{Serdyukov, Hiemstra, and Ruthven}{Serdyukov
  et~al\mbox{.}}{2011}]%
        {serdyukov2011towards}
\bibfield{author}{\bibinfo{person}{Pavel Serdyukov}, \bibinfo{person}{Djoerd
  Hiemstra}, {and} \bibinfo{person}{Ian Ruthven}.}
  \bibinfo{year}{2011}\natexlab{}.
\newblock \showarticletitle{Towards accessible search systems}. In
  \bibinfo{booktitle}{\emph{ACM SIGIR Forum}}, Vol.~\bibinfo{volume}{44}. ACM
  New York, NY, USA, \bibinfo{pages}{23--27}.
\newblock


\bibitem[\protect\citeauthoryear{Shaikh, Pera, and Ng}{Shaikh
  et~al\mbox{.}}{2015}]%
        {shaikh2015suggesting}
\bibfield{author}{\bibinfo{person}{Meher~T Shaikh},
  \bibinfo{person}{Maria~Soledad Pera}, {and} \bibinfo{person}{Yiu-Kai Ng}.}
  \bibinfo{year}{2015}\natexlab{}.
\newblock \showarticletitle{Suggesting simple and comprehensive queries to
  elementary-grade children}. In \bibinfo{booktitle}{\emph{2015 IEEE/WIC/ACM
  International Conference on Web Intelligence and Intelligent Agent Technology
  (WI-IAT)}}, Vol.~\bibinfo{volume}{1}. IEEE, \bibinfo{pages}{252--259}.
\newblock


\bibitem[\protect\citeauthoryear{Shivakumar and Georgiou}{Shivakumar and
  Georgiou}{2020}]%
        {shivakumar2020transfer}
\bibfield{author}{\bibinfo{person}{Prashanth~Gurunath Shivakumar} {and}
  \bibinfo{person}{Panayiotis Georgiou}.} \bibinfo{year}{2020}\natexlab{}.
\newblock \showarticletitle{Transfer learning from adult to children for speech
  recognition: Evaluation, analysis and recommendations}.
\newblock \bibinfo{journal}{\emph{Computer speech \& language}}
  \bibinfo{volume}{63} (\bibinfo{year}{2020}), \bibinfo{pages}{101077}.
\newblock


\bibitem[\protect\citeauthoryear{Unicef et~al\mbox{.}}{Unicef
  et~al\mbox{.}}{1989}]%
        {unicef1989convention}
\bibfield{author}{\bibinfo{person}{Unicef} {et~al\mbox{.}}}
  \bibinfo{year}{1989}\natexlab{}.
\newblock \showarticletitle{Convention on the Rights of the Child}.
\newblock  (\bibinfo{year}{1989}).
\newblock


\bibitem[\protect\citeauthoryear{Usta, Altingovde, Ozcan, and Ulusoy}{Usta
  et~al\mbox{.}}{2021}]%
        {usta2021learning}
\bibfield{author}{\bibinfo{person}{Arif Usta}, \bibinfo{person}{Ismail~Sengor
  Altingovde}, \bibinfo{person}{Rifat Ozcan}, {and} \bibinfo{person}{Ozgur
  Ulusoy}.} \bibinfo{year}{2021}\natexlab{}.
\newblock \showarticletitle{Learning to Rank for Educational Search Engines}.
\newblock \bibinfo{journal}{\emph{IEEE Transactions on Learning Technologies}}
  (\bibinfo{year}{2021}).
\newblock


\bibitem[\protect\citeauthoryear{Vanderschantz and Hinze}{Vanderschantz and
  Hinze}{2017a}]%
        {vanderschantz2017internet}
\bibfield{author}{\bibinfo{person}{Nicholas Vanderschantz} {and}
  \bibinfo{person}{Annika Hinze}.} \bibinfo{year}{2017}\natexlab{a}.
\newblock \showarticletitle{Do internet search engines support children's
  search query construction: a visual analysis}.
\newblock  (\bibinfo{year}{2017}).
\newblock


\bibitem[\protect\citeauthoryear{Vanderschantz and Hinze}{Vanderschantz and
  Hinze}{2017b}]%
        {vanderschantz2017study}
\bibfield{author}{\bibinfo{person}{Nicholas Vanderschantz} {and}
  \bibinfo{person}{Annika Hinze}.} \bibinfo{year}{2017}\natexlab{b}.
\newblock \showarticletitle{A Study of Children’s Search Query Formulation
  Habits}. In \bibinfo{booktitle}{\emph{Proceedings of the 31st International
  BCS Human Computer Interaction Conference (HCI 2017) 31}}.
  \bibinfo{pages}{1--4}.
\newblock


\bibitem[\protect\citeauthoryear{Vtyurina}{Vtyurina}{2019}]%
        {vtyurina2019towards}
\bibfield{author}{\bibinfo{person}{Alexandra Vtyurina}.}
  \bibinfo{year}{2019}\natexlab{}.
\newblock \showarticletitle{Towards Non-Visual Web Search}. In
  \bibinfo{booktitle}{\emph{Proceedings of the 2019 Conference on Human
  Information Interaction and Retrieval}}. \bibinfo{pages}{429--432}.
\newblock


\bibitem[\protect\citeauthoryear{Vtyurina, Fourney, Morris, Findlater, and
  White}{Vtyurina et~al\mbox{.}}{2019}]%
        {vtyurina2019verse}
\bibfield{author}{\bibinfo{person}{Alexandra Vtyurina}, \bibinfo{person}{Adam
  Fourney}, \bibinfo{person}{Meredith~Ringel Morris}, \bibinfo{person}{Leah
  Findlater}, {and} \bibinfo{person}{Ryen~W White}.}
  \bibinfo{year}{2019}\natexlab{}.
\newblock \showarticletitle{VERSE: Bridging screen readers and voice assistants
  for enhanced eyes-free web search}. In \bibinfo{booktitle}{\emph{The 21st
  International ACM SIGACCESS Conference on Computers and Accessibility}}.
  \bibinfo{pages}{414--426}.
\newblock


\bibitem[\protect\citeauthoryear{Wild}{Wild}{1996}]%
        {wild1996researching}
\bibfield{author}{\bibinfo{person}{Martyn Wild}.}
  \bibinfo{year}{1996}\natexlab{}.
\newblock \showarticletitle{Researching the use of the Internet with young
  children}. In \bibinfo{booktitle}{\emph{1995). Proceedings of Edith Cowan
  Memorial Conference}}. \bibinfo{pages}{41--50}.
\newblock


\bibitem[\protect\citeauthoryear{Yarosh, Thompson, Watson, Chase, Senthilkumar,
  Yuan, and Brush}{Yarosh et~al\mbox{.}}{2018}]%
        {yarosh2018children}
\bibfield{author}{\bibinfo{person}{Svetlana Yarosh}, \bibinfo{person}{Stryker
  Thompson}, \bibinfo{person}{Kathleen Watson}, \bibinfo{person}{Alice Chase},
  \bibinfo{person}{Ashwin Senthilkumar}, \bibinfo{person}{Ye Yuan}, {and}
  \bibinfo{person}{AJ~Bernheim Brush}.} \bibinfo{year}{2018}\natexlab{}.
\newblock \showarticletitle{Children asking questions: speech interface
  reformulations and personification preferences}. In
  \bibinfo{booktitle}{\emph{Proceedings of the 17th ACM Conference on
  Interaction Design and Children}}. \bibinfo{pages}{300--312}.
\newblock


\bibitem[\protect\citeauthoryear{Young, Hudry, Trembath, and Vivanti}{Young
  et~al\mbox{.}}{2016}]%
        {young2016children}
\bibfield{author}{\bibinfo{person}{Nicole Young}, \bibinfo{person}{Kristelle
  Hudry}, \bibinfo{person}{David Trembath}, {and} \bibinfo{person}{Giacomo
  Vivanti}.} \bibinfo{year}{2016}\natexlab{}.
\newblock \showarticletitle{Children with autism show reduced information
  seeking when learning new tasks}.
\newblock \bibinfo{journal}{\emph{American journal on intellectual and
  developmental disabilities}} \bibinfo{volume}{121}, \bibinfo{number}{1}
  (\bibinfo{year}{2016}), \bibinfo{pages}{65--73}.
\newblock


\end{thebibliography}

\end{document}